\documentclass[12pt]{article}

\def\ltsim{\lower3pt\hbox{$\, \buildrel < \over \sim \, $}}
\def\gtsim{\lower3pt\hbox{$\, \buildrel > \over \sim \, $}}
\def\be{\begin{equation}}
\def\ee{\end{equation}}
\def\ba{\begin{eqnarray}}
\def\ea{\end{eqnarray}}
\def\ga{\mathrel{\raise.3ex\hbox{$>$\kern-.75em\lower1ex\hbox{$\sim$}}}}
\def\la{\mathrel{\raise.3ex\hbox{$<$\kern-.75em\lower1ex\hbox{$\sim$}}}}

\def\bo{ { \sqcup\llap{ $\sqcap$} } }
\begin{document}

\baselineskip=16pt 
\begin{titlepage}
\rightline{OUTP-01-04P}
\rightline{hep-th/0102015}
\rightline{February 2001}  
\begin{center}

\vspace{2cm}

\large {\bf Dilaton tadpoles and mass in warped models}

\vspace*{15mm}
\normalsize

{\bf  Antonios Papazoglou\footnote{a.papazoglou@physics.ox.ac.uk}}

\smallskip 
\medskip 
{\it Theoretical Physics, Department of Physics, Oxford University}\\
{\it 1 Keble Road, Oxford, OX1 3NP,  UK}
\smallskip

\vskip0.6in \end{center}
 
\centerline{\large\bf Abstract}

We review the brane world sum rules of Gibbons at al for compact five
dimensional warped models with identical four-geometries and bulk dynamics involving scalar
fields with generic potential. We show that the
absence of dilaton tadpoles in the action functional of the theory  is
linked to one of these sum rules. Moreover, we calculate the dilaton mass term
and  derive the condition that is necessary for stabilizing the system.

\vspace*{2mm} 

\end{titlepage}

\section{Introduction}

Recently Gibbons, Kallosh and Linde \cite{gibb} derived an
infinite set of sum rules for five dimensional models with a compact
periodic extra dimension and identical four geometries. These constraints were an immediate
consequence of the equations of motion and served as consistency
checks of several recent constructions. An interesting result was that the
Goldberger-Wise (GW) mechanism \cite{gw} of stabilizing the two three-brane
Randall-Sundrum (RS1) model \cite{RS1} has 
to include the backreaction on the metric in order to agree with a
specific constraint, something done in the
DeWolfe-Freedman-Gubser-Karch mechanism \cite{karch}. 

A particular sum rule that attracted the attention of \cite{gibb} was
the $\displaystyle{\oint dy~W (T_{\mu}^{\mu}-2T_5^5)=0}$, where $W(y)^2$ stands for
the warp factor. This constraint was firstly derived by \cite{ell} as a 
condition of vanishing of the four dimensional cosmological
constant. The interesting point was that this combination of the energy-momentum tensor components
appeared in a condition for the absence of dilaton\footnote{Here we
use the term ``dilaton''  to 
denote the modulus corresponding to the fluctuation of the overall
size of the system. We use instead the term ``radions'' for the
moduli associated with the position of freely moving branes  along the extra
dimension (not on
orbifold fixed points).}  tadpoles  in
the action functional of the theory in the paper by Kanti, Kogan, Olive and
Pospelov  \cite{kanti}. The condition of \cite{kanti} reads $\displaystyle{\oint dy\sqrt{G^{(5)}} (T_{\mu}^{\mu}-2T_5^5)=0}$. It was however pointed
out by \cite{gibb} that the two constraints were not identical
because the  $(T_{\mu}^{\mu}-2T_5^5)$ combination was integrated with
different powers of the warp factor since
$\sqrt{G^{(5)}}=W^4\sqrt{g}$.  A closer inspection reveals that these two constraints are actually
identical given the assumptions made in \cite{kanti}. In more detail,
the condition in \cite{kanti} was derived for matter dominated
branes were the effect of the warp factor is negligible. Then, for
$W\approx 1$ the two constraints coincide.

In this paper we will iterate the calculation of \cite{kanti},
including the full effect of the warp factor. In that case, the 
condition of \cite{kanti} for the absence of dilaton tadpoles is
modified and the new condition coincides with an other sum rule of
\cite{gibb}. Furthermore, having the quadratic action functional for the dilaton,
it is straightforward to read off its mass. We find a generic formula
relating the dilaton mass with the sum of the tensions of the
branes and the curvature of the four-geometries. Demanding that this
mass is not tachyonic we can derive the necessary condition for
stabilizing the overall size of the system. This is in accordance 
with the result found in \cite{gibb} that the GW stabilization
mechanism of the  RS1 model has 
to include the backreaction on the metric.

\section{Review of sum rules}

At first it would be instructive to review the sum rules presented in
\cite{gibb}. We will concentrate  as in \cite{gibb} in the case  where 
the background metric of the five
dimensional spacetime can be written in the form:
\be
ds^2=W(y)^2g_{\mu\nu}(x)dx^{\mu}dx^{\nu}+dy^2
\label{back}
\ee
with $g_{\mu\nu}(x)$ a general background four dimensional metric
and $W(y)$ a generic warp factor. We should stress here that this is
not the most general choice of metric in five dimensions as we have
explicitly assumed that the all four dimensional sections have the
same geometry.

We can now consider an arbitrary number of minimally coupled scalar bulk fields
$\Phi^I(x,y)$ with internal metric ${\mathcal{G}}_{IJ}$ and arbitrary
bulk potential $V(\Phi)$ (which includes bulk cosmological constant), 
coupled to an again arbitrary number of branes with brane potential
$\lambda_{i}(\Phi)$ (which again includes the brane tensions). The
action describing the above system is the following:
\be
S = \int d^{4}x dy \sqrt{-G}\left(2M^{3}R
-{1 \over
2}{\mathcal{G}}_{IJ}\partial_M\Phi^I\partial^M\Phi^J-V(\Phi)-\sum_{i}\lambda_{i}(\Phi)\delta(y-y_i){\sqrt{-\hat{G}^{(i)}} 
\over\sqrt{-G}} \right)
\label{action}
\ee
where $\hat{G}^{(i)}_{\mu\nu}$ is the induced metric on the brane and
$M$ the fundamental 5D scale. The
Einstein equations arising from the above metric can be written in
the form:
\ba
4M^3R_{\mu}^{\mu}&=&-{1 \over 3}T_{\mu}^{\mu}-{4 \over 3}T_5^5\label{e1}\\
4M^3R_5^5&=&-{1 \over 3}T_{\mu}^{\mu}+{2 \over 3}T_5^5\label{e2}
\ea
where the energy-momentum tensor components are:
\ba
T_{\mu}^{\mu}&=&-\partial_{\mu}\Phi\cdot\partial^{\mu}\Phi-2\Phi'\cdot\Phi'-4V(\Phi)-4\sum_{i}\lambda_{i}(\Phi)\delta(y-y_i)\label{T}\\
T_5^5&=&-{1 \over 2}\partial_{\mu}\Phi\cdot\partial^{\mu}\Phi+{1 \over 2}\Phi'\cdot\Phi'-V(\Phi)\label{T5}
\ea
with the indices in the above formulas raised and lowered by
$G_{\mu\nu}=W(y)^2g_{\mu\nu}(x)$ and where dot product denotes
construction with the internal metric ${\mathcal{G}}_{IJ}$. Since we are interested on a
background configuration, the $\partial_{\mu}\Phi\cdot\partial^{\mu}\Phi$ 
terms can be dropped. The Ricci tensor is easily calculated to be:
\ba
R_{\mu}^{\mu}&=&W^{-2}R_g -12W'^2W^{-2}-4W''W^{-1}\label{r}\\
R_5^5&=&-4W''W^{-1}\label{r5}
\ea

If we now consider the function $(W^a)''$ with $a$ an arbitrary real
number, its integral around the compact extra dimension is zero. Using
(\ref{e1}), (\ref{e2}), (\ref{r}), (\ref{r5}) we arrive at an infinite 
number of constraints \cite{gibb}:

\be
\oint dy~W^a(T_{\mu}^{\mu}+(2a-4)T_5^5)=4M^3(1-a)R_g\oint dy~W^{a-2}
\ee

As it is obvious, these constraints are a natural consequence of the
equations of motion. It is straightforward to see for example  that they are
satisfied in the RS1 model \cite{RS1} and the
bigravity/multigravity models \cite{multi,++}.  We will
single out three constraints which we will be important for the
subsequent discussion, namely the ones for $a=0,1,2$:
\ba
\oint dy~(T_{\mu}^{\mu}-4T_5^5)&=&4M^3R_g\oint dy~ W^{-2}\label{0gib}\\
\oint dy~W(T_{\mu}^{\mu}-2T_5^5)&=&0\label{gib}\\
\oint dy~W^2T_{\mu}^{\mu}&=&-4M^3R_g\oint dy\label{newgib}
\ea

One could also use the above constraints for non-compact models, but 
should be careful that the above derivation makes sense. For the
finite volume flat one three-brane Randall-Sundrum model (RS2)
\cite{RS2}, all constraints are valid for $a\geq 0$. For the infinite volume
Gregory-Rubakov-Sibiryakov model (GRS) \cite{GRS}  only the $a=0$ constraint is
valid and for the infinite volume Karch-Randall model (KR) \cite{KR} all constraints
are valid for $a\leq 0$ .

\section{Dilaton in warped backgrounds}

We now consider the perturbation related to the overall size of the
compact system, namely the dilaton. The general form of the metric for 
the physical radion perturbations that do not mix with the graviton(s)
is given in \cite{pilo}. For the dilaton, the ansatz is rather simple and can be
written in the form \cite{ruth} (see also \cite{bagger}):
\be
ds^2=e^{-W(y)^{-2}\gamma(x)}W(y)^2g_{\mu\nu}(x)dx^{\mu}dx^{\nu}+\left(1+W(y)^{-2}\gamma(x)\right)^2dy^2\label{ruth}
\ee

Substituting the above metric in the action (\ref{action}) (see
Appendix for analytic formulas),
integrating out total derivatives, throwing out $\gamma$-independent parts  and keeping terms up to quadratic order, we get:
\be
S= \int d^4xdy\sqrt{g}\left\{-{1 \over
2}(6M^3W^{-2})g^{\mu\nu}\gamma_{,\mu}\gamma_{,\nu}+{\mathcal{L}}_1\gamma-{1 
\over 2}{\mathcal{L}}_2\gamma^2\right\}
\ee
with
\ba
{\mathcal{L}}_1&=&2M^3(4W'^2+16W''W)+{3 \over
2}W^2\Phi'\cdot\Phi'+W^2V(\Phi)+2W^2\sum_{i}\lambda_{i}(\Phi)\delta_(y-y_i)~~~~~~~~~~~\\
{\mathcal{L}}_2&=&2M^3(W^{-2}R_g-32W'^2W^{-2}+32W''W^{-1})+5\Phi'\cdot\Phi'+4\sum_{i}\lambda_{i}(\Phi)\delta_(y-y_i)
\ea


At this point, let us work out the integral over the extra dimension
of the tadpole term ${\mathcal{L}}_1$  of the Lagrangian. This gives:
\be
\oint dy{\mathcal{L}}_1=-{1 \over 2}\oint dy~W^2\left[T_{\mu}^{\mu}-2T_5^5-16M^3\left({W'^2 \over W^2}+4{W'' \over W}\right)\right]\label{improved}
\ee
where we used the energy-momentum tensor components found in the previous 
section with respect to the unperturbed background metric (\ref{T}),
(\ref{T5}). We can further simplify this quantity if we use the
equations (\ref{e1}), (\ref{e2}), (\ref{r}), (\ref{r5}) which hold for 
the background metric. The resulting expression is:
\be
\oint dy{\mathcal{L}}_1={1 \over 6}\oint dy~W^2\left[T_{\mu}^{\mu}+4M^3W^{-2}R_g\right]
\ee
which is exactly zero because of the $a=2$ constraint
(\ref{newgib}). This result should have been expected since the
perturbation (\ref{ruth}) is bound to extremize the effective
potential when one evaluates the action using the
background equations of motion. However, it is interesting and rather
unexpected that the absence of the tadpole term is linked to this particular sum rule of \cite{gibb}.  It is worth
mentioning here that in the case that the warp factor is effectively
constant ($W\approx 1$), as it was assumed in \cite{kanti}, the condition that the expression
(\ref{improved}) vanishes, is identical with the   $a=1$ constraint (\ref{gib}).

Our next task is to read off the mass of the dilaton from the action
functional. For this reason we define the canonically normalized
dilaton field  with mass dimension one $\bar{\gamma}^2=\left(6M^3\oint
dyW^{-2}\right)\gamma^2\equiv A \gamma^2$. Then the
mass of the canonical dilaton $\bar{\gamma}$ is:
\be
m^2={1 \over A}\oint dy{\mathcal{L}}_2
\ee
After a lot of simplifications using the relations
(\ref{e1})-(\ref{r5}) we obtain:
\be
m^2=-{1 \over 3A}\oint dy(10M^3W^{-2}R_g +\Phi'\cdot\Phi'+4\sum_{i}\lambda_{i}(\Phi)\delta(y-y_i))
\ee

We can further simplify the expression using the $a=0$ constraint
(\ref{0gib}) and get a more suggestive result:
\be
m^2={1 \over A}\oint dy(\Phi'\cdot\Phi'-2M^3W^{-2}R_g)\label{fi}
\ee
or equivalently,
\be
m^2=-{1 \over A}\left\{\sum_{i}\lambda_{i}(\Phi)+3M^3R_g\oint dyW^{-2}\right\}\label{V}
\ee

From the second expression it is clear that we cannot have a massive
dilaton if the sum of the effective tensions of the branes
$\lambda_{i}(\Phi)$ is exactly zero and at the same time they are
kept flat. This is the same conclusion that appeared in \cite{gibb}
regarding the GW mechanism in the RS1
scenario. Moreover, the absence of tachyonic mass would guarantee the
stabilization of the overall size of any system with the above
characteristics. By eqs.(\ref{fi}),(\ref{V}) we get two equivalent conditions:
\ba
\oint dy(\Phi'\cdot\Phi'-2M^3W^{-2}R_g) &>& 0\\
\sum_{i}\lambda_{i}(\Phi)+3M^3R_g\oint dyW^{-2} &<& 0
\ea

If one wishes to have flat branes, then  the sum of brane
tensions should be negative or equivalently one should have a non constant (in $y$) scalar field
configuration. In the case that the above expressions (and thus the mass)
vanish, one should look for the higher orders of the effective
potential to examine the stability of the system.

Finally, we should note that the formulas (\ref{fi}), (\ref{V}) are valid even
for non-compact models whenever the dilaton mode is normalizable. This 
happens for example in the KR model \cite{KR} and one can find from the above
expressions the mass of the dilaton, in agreement with \cite{adsrad}. In
that case, the dilaton mode cannot be attributed to the fluctuation of
the overall size of the system, but can be understood to be a remnant
mode if we start with the compact $''++''$ system \cite{++} and send
one of the branes to infinity (\textit{i.e.} after decompactifying the system). 

\section{Conclusions}

We have showed that in a general warped metric with identical
four-geometries and arbitrary bulk dynamics involving minimally
coupled scalar fields, the absence of dilaton tadpoles is related to
one particular sum rule of \cite{gibb}. Moreover, we have
calculated the dilaton mass as a function of the sum of the brane
tensions and the leftover curvature of the branes. The result agrees
with the observation made by \cite{gibb} that one could not have a
massive dilaton for flat four-geometries and zero net brane tensions. 

It would be interesting to see what happens with higher than
quadratic terms in the dilaton potential and the possible role that
the other sum rules of \cite{gibb} play. Moreover, one could work out
the same calculation for the other moduli in these configurations, the
radions \cite{pilo}, and see if/how the results are modified. Finally, a much more
general investigation is needed to obtain the sum rules and their
role for the dilaton and radion potentials in models in which the four 
dimensional geometries are not identical as it happens with
cosmological solutions (see \textit{e.g.} \cite{cosmol}). These are important issues for understanding
the dynamics of the dilaton/radions in the extra dimensional models and 
will be addressed in an other publication \cite{future}.

\textbf{Acknowledgments:} We would like to thank Gary Gibbons, Stavros 
Mouslopoulos, Luigi Pilo and Graham G. Ross for helpful
discussions. We are grateful to Ian I. Kogan and Keith A. Olive for
valuable discussions. We are 
indebted to Panagiota Kanti for important comments and careful
reading of the manuscript. This work is supported by the Hellenic State Scholarship Foundation (IKY) \mbox{No. 8017711802}.

\def\theequation{A.\arabic{equation}}
\setcounter{equation}{0}
\vskip0.8cm
\noindent
\centerline{\Large \bf Appendix}
\vskip0.4cm
\noindent

In this appendix we list the Ricci tensor components, the Ricci
scalar and the action obtained by the metric:
\be
ds^2=e^{-W(y)^{-2}\gamma(x)}W(y)^2g_{\mu\nu}(x)dx^{\mu}dx^{\nu}+\left(1+W(y)^{-2}\gamma(x)\right)^2dy^2
\ee
The spacetime components of the five dimensional Ricci tensor are:
\ba
R_{\mu\nu}~=~R_{g~\mu\nu}&+&{g_{\mu\nu} \over 2}W^{-2}\bo\gamma -{g_{\mu\nu}
\over 2}{W^{-6}\gamma \over
(1+W^{-2}\gamma)}\gamma_{,\kappa}\gamma^{,\kappa}-{1 \over 2}\left({1-W^{-2}\gamma \over
1+W^{-2}\gamma}\right)W^{-4}\gamma_{,\mu}\gamma_{,\nu}\nonumber\\&+&{W^{-4}\gamma \over
(1+W^{-2}\gamma)}D_{\mu}\partial_{\nu}\gamma- g_{\mu\nu}
e^{-W^{-2}\gamma}W'^2 \left({3+4W^{-2}\gamma \over
1+W^{-2}\gamma}\right)\nonumber\\&-&g_{\mu\nu}e^{-W^{-2}\gamma}{ WW'' \over (1+W^{-2}\gamma)}
\ea
and the (55) component:
\ba
R_{55}&=&e^{W^{-2}\gamma}(1+W^{-2}\gamma)W^{-6}\gamma_{,\mu}\gamma^{,\mu}-e^{W^{-2}\gamma}(1+W^{-2}\gamma)W^{-4}\bo\gamma\nonumber\\&\phantom{=}&-4(1+W^{-2}\gamma)W^{-1}W''-4(1+W^{-2}\gamma)W^{-4}W'^2\gamma
\ea
Finally the Ricci scalar is:
\ba
R&=&e^{W^{-2}\gamma}W^{-2}R_g+e^{W^{-2}\gamma}\left({1+3W^{-2}\gamma \over
1+W^{-2}\gamma}\right)W^{-4}\bo\gamma+{1 \over 2}e^{W^{-2}\gamma}\left({1-3W^{-2}\gamma \over
1+W^{-2}\gamma}\right)W^{-6}\gamma_{,\mu}\gamma^{,\mu}\nonumber\\&\phantom{=}&-8{
W^{-1}W'' \over (1+W^{-2}\gamma)}-{12W^{-2}W'^2+20W^{-4}W'^2\gamma
\over(1+W^{-2}\gamma) }
\ea
In the above expressions the indices are raised and lowered by
$g_{\mu\nu}$.

The action (\ref{action}) then becomes:
\ba
S&=&\int d^4xdy \sqrt{g}
\left\{2M^3\left[e^{-W^{-2}\gamma}W^2(1+W^{-2}\gamma)R_g+e^{-W^{-2}\gamma} 
\left(1+{5\over 2}W^{-2}\gamma +{3 \over
2}W^{-4}\gamma^2\right)g^{\mu\nu}D_{\mu}\partial_{\nu} \gamma
\right.\phantom{\sum_{i}}
\right.\nonumber\\&\phantom{=}&~~~~~~~~~~~~~~~~~~~~~~~~~~~~~~~~~~~~~~~~~~~~~~~~~~~~\phantom{5 
\over 2}\left.+e^{-2W^{-2}\gamma}(-8W''W^3-12W'^2W^2-20W'^2\gamma)\right]\nonumber\\&\phantom{=}&~~~~~~~~~~~~-{1 
\over 2}{W^4 \over
(1+W^{-2}\gamma)}e^{-2W^{-2}\gamma}\Phi'\cdot\Phi'-e^{-2W^{-2}\gamma}W^4(1+W^{-2}\gamma)V(\Phi)\nonumber\\&\phantom{=}&~~~~~~~~~~~~\left.-e^{-2W^{-2}\gamma}W^4\sum_{i}\lambda_{i}(\Phi)\delta_(y-y_i)\right\}
\ea


\end{document}